# Collaborative process control: Observation of tracks generated by PLM system

S. El Kadiri, P. Pernelle, M. Delattre, A. Bouras
*LIESP – Université de Lyon – France*
[soumaya.el-kadiri] [miguel.delattre] [abdelaziz.bouras]@univ-lyon2.fr,
philippe.pernelle@iutb.univ-lyon1.fr

***Abstract:*** *This paper aims at analyzing the problems related to collaborative work using a PLM system. This research is mainly focused on the organisational aspects of SMEs involved in networks composed of large companies, subcontractors and other industrial partners. From this analysis, we propose the deployment of an approach based on an observation process of tracks generated by PLM system. The specific contributions are of two fold. First is to identify the brake points of collaborative work. The second, thanks to the exploitation of generated tracks, it allows reducing risks by reacting in real time to the incidents or dysfunctions that may occur. The overall system architecture based on services technology and supporting the proposed approach is described, as well as associated prototype developed using an industrial PLM system.*

# 1. INTRODUCTION

Most of industrial activities take place within collaboration with multiples partners (subcontractor, customers …). At the same time this situation leads to many brakes that require, among others, the use of regulation mechanisms and tools (Ithnin N. & al 2006).
One of the goals of PLM reducing the product life-cycle is to foster collaboration among different actors involved in business processes of the product lifecycle (design, development; production; sale; retrieve) (Debaecker.D. 2004). Thanks to the standard tools in the PLM systems (forum, workflow, messaging…), users have the means to support collaborative projects.
Nevertheless, in a more and more competitive environment constraining organization to a constant research of balance, the collaborative processes remain marked by their instability and by the incompleteness and a relative incoherence of the rules that govern them (Green P. & al 2000).
We quote for example a diversion processes; a rigidity of the work practices that is in contradiction with the circulation of information and the requested rapidity of processes; sometimes, the loss of actors' autonomy in the process of communication; the ambivalence linked to the production of tracks and to their exploitation; the incompleteness of the work practices descriptions (Mishra A. & al 2006). The SME more particularly remain reluctant to implement such systems (Pol G. & al 2007).

The brakes analysis of the traditional schema of collaboration within PLM systems points out a lack of agility on processes; justified by the willingness to control all business processes. This drives the lead team designs processes mobilizing many product data (Morley C. 2004). It results in increasing work load by multiplying the validation tasks, decreasing reactive capabilities of actors regarding the customer needs, etc. The lack of agility is situated also within organizational structures. In fact, the management of access rights related to organizational levels is often too constraining and does not always allow a sufficiently global vision of the users access capacities, inducing sometimes the loss of autonomy of principal actors in the process (El Kadiri S. & al 2008a).

To synthesize, the obstacles to an effective collaborative work related to product lifecycle management are mostly related to the level of control: "the control level within processes and organization seems conditioning collaboration quality".

## 2. RESEARCH METHODOLOGY

The CSCW researches propose a methodological framework to facilitate collaborative work among a group of actors. One of the aspects covered by the CSCW is the interactions analysis between actors which led to collaborative activities segmentation. Moreover this work has resulted in the classical distinction between various activities: communication, coordination, collaboration. This segmentation is incomplete in the context of an industrial activity leading to new products development. Indeed, it is advisable to have an organizational approach regarding to constraints induced on the collaboration modes and tools (notably processes).
In this respect, we propose two ways of analyzing the problem.

A first analyse stage is related to the **organizational aspect**: the consideration of the organizational context is an essential element of differentiation (Delattre M. 2001). Indeed, according to the industrial structure (international groups, networked enterprise, sub-contracting, design outsourcing, etc) collaborations between actors are subjected to different constraints. We can, for instance, identify:
- *Hierarchical constraints*, which typically impose restrictions if people are not on the same level.
- *Functional constraints*, established by the enterprise or by the external constraints, such as specific certification or rules related to a specific sector (chemistry, food processing industry) that imposes to respect some procedures.
- *Communautarist constraints*, which are more diffuse but always present in all areas and whatever activity, inducing by the way the feeling of membership to a group of actors rather than to another.
- *Customer constraints* depend on the degree of his involvement in the company's activity. In the case of outsourcing, the customer is very active and collaborative activities are generally determined by the following paradox: satisfying the customer by providing him the maximum of information without delivering the expertise that maintains the company's sustainability.

The incidence of these constraints affects partially, but in explicit way, the PLM system. For instance, the hierarchical constraints affect the management of access rights to the objects; and the functional constraints are implemented in the PLM workflows. The collaborative problems come in fact from these constraints that, implicitly or indirectly, affect the collaborative activity.

In this respect, and in order to reveal these constraints, to analyze the consequences of these ones to propose actions in order to minimize their scope and impacts, a first approach consists in the development of a ***process of observation*** of the collaborative activity.

The work in the ILE (Interactive Learning Environment) fields and especially the approach by the agents (Marty J-C. & al. 2007) identifies several types of observation agents: collector agent, historic agent, structuring agent, statistical agent, etc.

In the context of PLM systems, we propose to use the following concepts:
- *Collector agent* is responsible for collecting traces from user actions.
- *Structuring agent* is specialized for grouping, reformatting or annotating the collected traces, to make them more suitable for the construction of indicators.
- *Statistical agent* is responsible for establishing statistics on the objects' usage.
- *Notification agent is specialized for notifying the regulator basing on predefined alert levels and threshold triggers.*
- *Visualisation agent allows representing the measured indicators via the statistical agent through dashboards.*

The traceability capabilities proposed by PLM systems allows identifying many traces left by the actor's activity within the system. The observation of collaborative activities using the different agents will allow us to identify different data participating to most activities. The exploitation of these ones will be based on the construction of monitoring indicators.

The second stage analysis treats the processes. The collaboration within PLM systems is essentially managed by business processes. The work methods (processes) cannot be fixed definitively (Berthier D. 2006), in the sense that they depend on the collaboration context; any state change within the first is liable to a state change within the latter.

Several parameters condition the collaboration activity; and the analysis of the reality of the first is inseparable of the latter. These parameters must be identified and integrated to the PLM system in order to aim progressively for flexible (and agile) processes (Barrand J. 2006). To illustrate these matters, one of the parameters that we place to the heart of the collaborative work is the "confidence relation". The access rights to the PLM system attributed to a customer vary according to the level of the evaluated confidence. The evaluation of this one can result from the observation process described in the 1st stage analysis.

In this paper, we focus on the first stage analysis, consisting in observation process development.

## 3. METHODOLOGY AND DATA

### 3.1. Observation of collaborative activities

We wish observe how the collaborative activity unfolds. Why is this observation indispensable? First of all, the observation aims to the regulation of the activity. It allows obtaining useful information for the adapted sequence of collaborative activity continuation. The observation produces too much useful information (Carron T. & al 2006). The activity quality can be evaluated in the same manner than software processes, for example with the CMM model. The idea is to modify and to enrich the scenario in which certain activities are for instance systematically added or eliminated by the users.

The approaches to accomplish this observation are multiple (Gligor-Calin L. 2006): quality techniques based on interviews or questionnaires regarding to the use of a specific system; quantitative techniques based notably on the analysis of the server (logs, data base, and application server), the client (man-machine interactions), etc.
Concerning quantitative techniques, the analysis of the distinct observation sources is necessary in order to identify data participating to the activity to be managed. To this effect, a track analysis is a posteriori necessity to identify redundant, missing and useless steps. Two sources of observations can be used: the server and the client.
Thus, each of the different sources necessitates a particular instrumentation in order to help the interpretation of collected data (Heraud J-M. & al 2005). In this paper we focus at first on the exploitation of the log files.
Log files contain all actions performed on the server. The observed elements are not directly exploitable. The tracks generation from the log files is a complicated process that needs many operations: cleaning, annotation of the raw observations …

A first approach incites first of all to orchestrate the PLM system in order to observe abstract elements; and therefore to proceed to their analyze by the bias of monitoring indicators (Time spent on a given task, number of refusal, etc.). Afterward, dashboards will be developed. A second approach consists on the deployment of a heuristic based on the indicators previously measured. This heuristic will proceed to the notification of problems and detected malfunctions.

### 3.2. Observation within PLM system

PLM systems integrate a set of meta-models (product data, process and organization). Each instanced class of theses meta-models is involved in the product development. However, all these elements do not contribute in the same way to the collaborative process. From the meta-data and processes introduced, we propose to identify present objects (class document, specific CAD model…) and flow control in this collaborative process.

Hence, we describe the process of abstract elements generation from logs as follow:

*1- Collection of information (via the collector agent):*
It consists on selecting from the different log files and preparing data that will allow extracting information concerning the different conducted actions.

*2- Structuring collected information (via the structuring agent):*
It consists on transforming these data in an *extraction context*. This latter represents a triplet constituted of the following parameters: activity, object, and actor.
- Activity: creation, updating, deleting, link management, statute management, locking, visualization, index management, information research.
- Object: document, CAO models, sets, form, composition piece, process models.
- Actor, the system users: internal and external participant to the activity.

*3- Identification of collaborative data (via the statistical agent):*
It aims on the identification of frequent triplets through a measure function, and a threshold. This function takes in parameter the identified triplets at the previous step. The measure can be the occurrence number, the length, the frequency, etc.
For instance, the measure function may be: frequency of a given object, number of modification, number of access by more two actors, number of task where the object is in output of flow…

## *3.2. Exploitation of observation results*

The generation process is followed by the extracted tracks exploitation process. It is described as follows:

*4- Measuring of monitoring indicators (via the statistical agent):*
The implementation of monitoring indicators provides a performance management and monitoring of collaborative processes. Indeed, the identification of problems relies on the interpretation of monitoring indicators supplied by the agents of the observation.
The construction of the indicators was realized in one of our previous research works (El Kadiri S. & al, 2008b). We quote some examples of indicators:
- Number of refused change/validation requests (IP2- C1.1);
- Number of non respected times to realize a given task(IP11- C4.1);
- Time spent on information search on a same object (reactive capacity of the actors) (IP7- C2.1.1, C3.1.5);
- Number of changes performed on a specific process model (IP4- C1.3.2).

*5- Notification of the regulator (via the notification agent):*
The regulator must define its own alter levels and triggering thresholds. Thus he will be notified depending on these parameters.

*6- Visualisation (via the visualisation agent):*
It consists on the construction of dashboards. It offers to the regulator mechanisms in order to support the decision making.

The two processes are illustrated in the following figure.

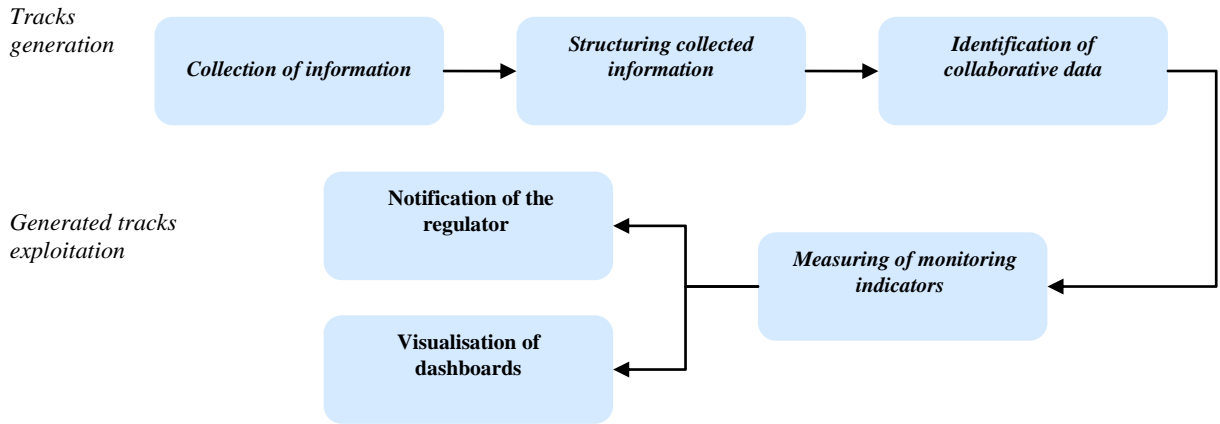

**Figure 1 :** *Tracks generation and exploitation processes*

The meta-models supporting this processes was described in one of our previous works "The extended meta-model" in (El Kadiri S. & al 2008a).

The following figure highlights the interaction between the class model and the different agent concepts.

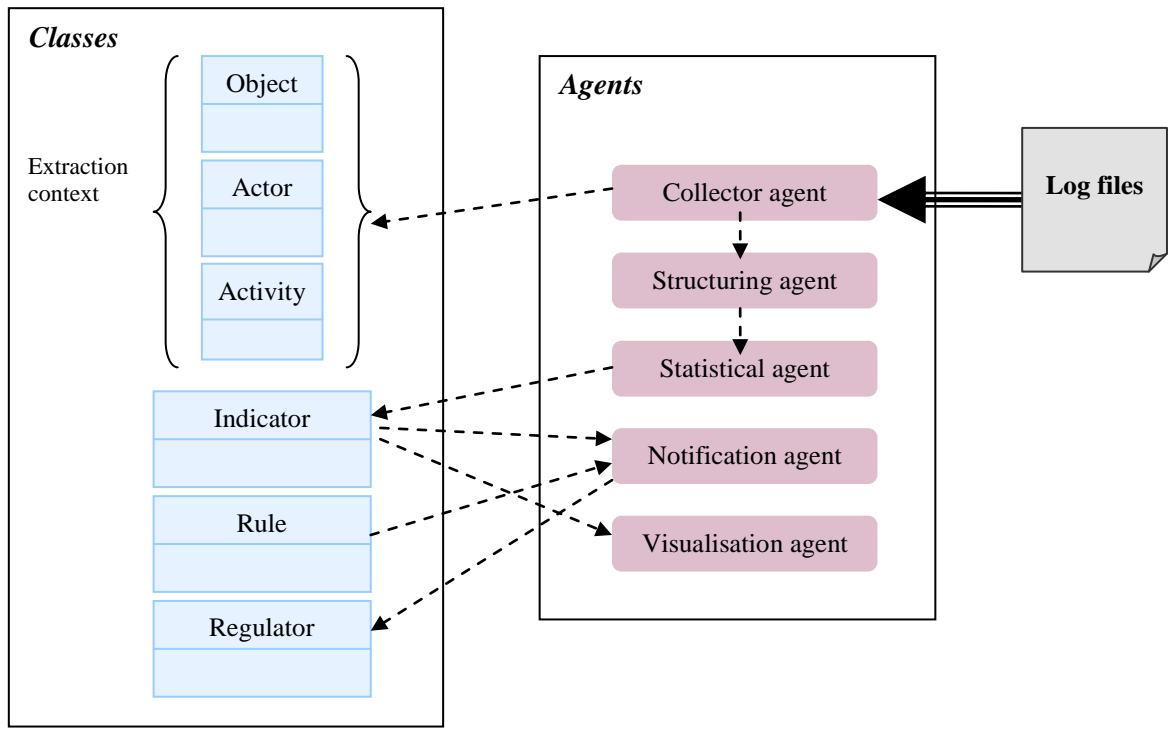

**Figure 2:** *Interaction Classes and Agents*

# 4. THE PROPOSED ARCHITECTURE

## 4.1. The architecture

In order to implement the described approach, we propose to develop a standardized layer of configurable services to connect to PLM system by the bias of interfaces. These services offer the described functionalities within the tracks extraction process and generated tracks exploitation process.

This service layer leans on architecture three third:
- presentation layer: corresponds to the man/machine interface;
- business layer : corresponds to the to the implementation of the logical applicative;
- data access layer: aims to stock the collected data through the collector agent.

We consider three observation sources may be used in order to perform the observation process.
- application server,
- data base server,
- log files.

At this stage, only the two last sources are deployed. The resulted architecture is illustrated in the following figure (Figure 3).

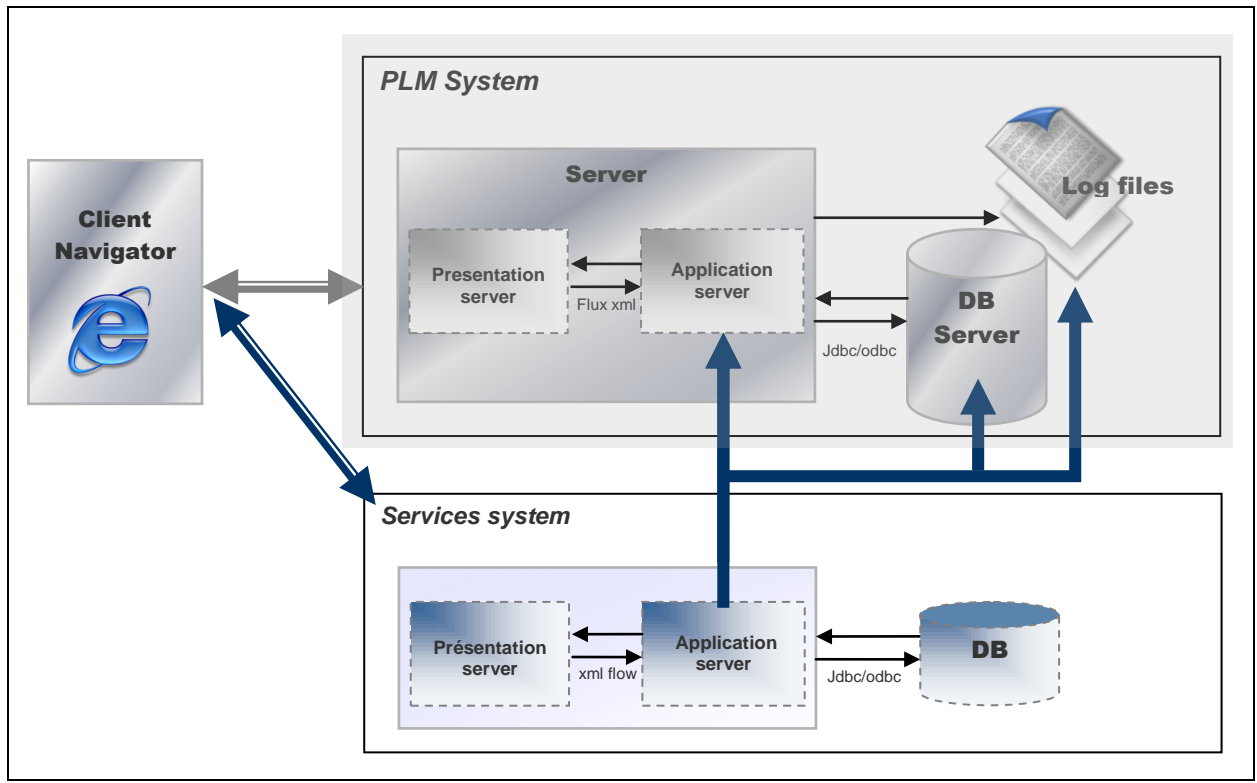

**Figure 3 :** *The proposed system architecture*

## 4.2. Deployment

We use the server as an observation source (data base and log files). Our experimentation platform @udros use the logging Java API: ***Log4j***. It is a highly configurable system, facilitating thus the observation thanks to the generated logs that contain all actions performed on @udros. We have already developed the associated services, which represent Java programs connected to the @udros system. The following figure (Figure 4) illustrates this implementation.

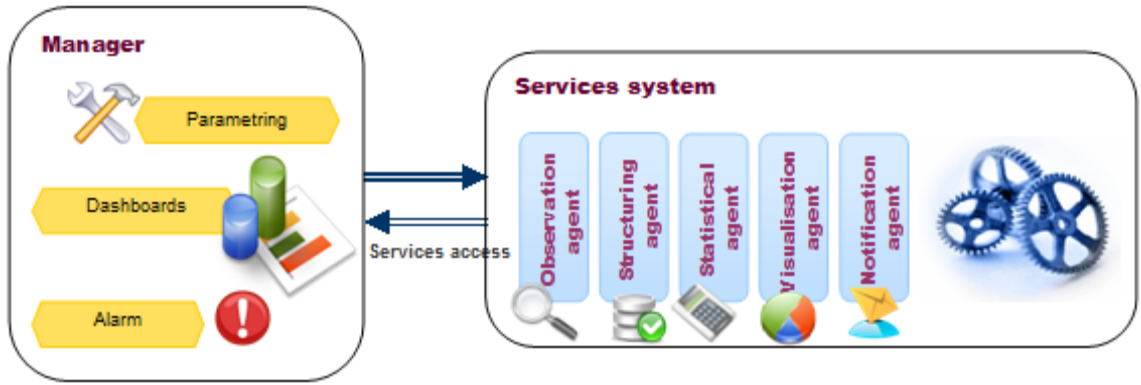

**Figure 4:** *Implementation with the PLM system @udros*

To validate this implementation, we implement a prototype with some indicators. The developed dashboards are illustrated on Figure 5. The used indicators are:
- Number of activities performed by actors (1$^{st}$ dashboard on left)
- Percentage of activities performed on objects (2$^{nd}$ dashboard on left)
- Number of changes performed on a specific process model (dashboard on right)

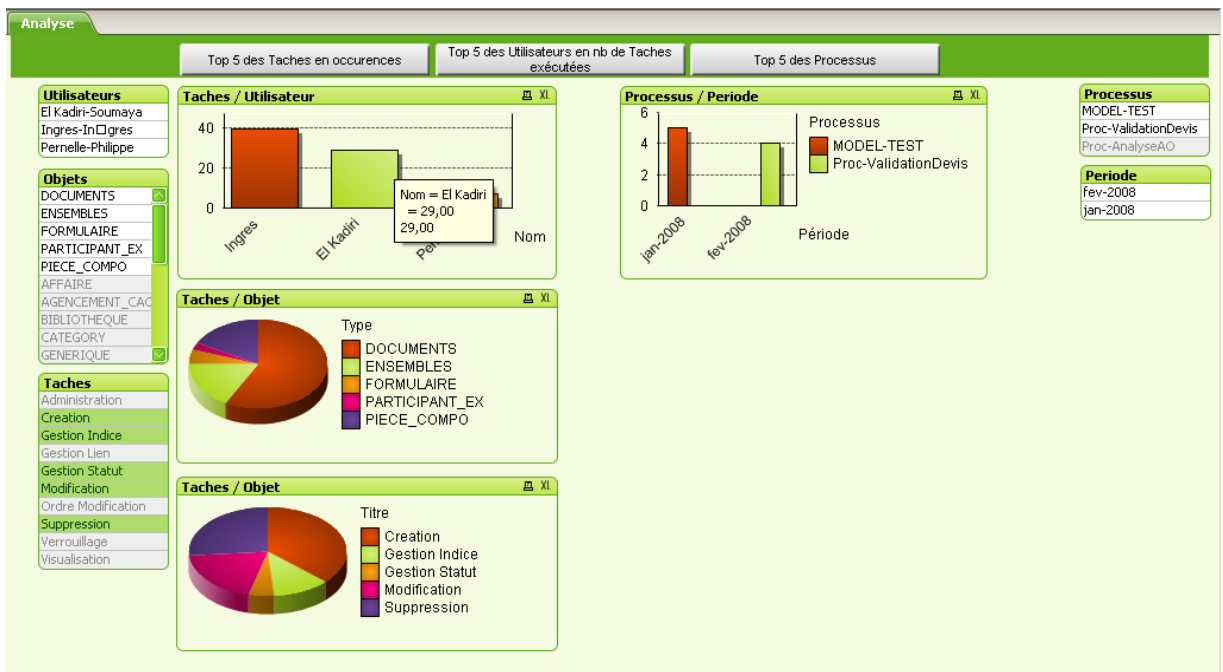

**Figure 5:** *The developed **p**rototype – Example of dashboards*

We expected to experiment this services layer on an industrial case in the plastics industry. Unfortunately, the enterprise is still implementing the PLM system. The realized experimentation is limited at this stage of our works to the validation of the prototype.

## 5. CONCLUSION

This paper discussed the problems associated with the SMEs collaboration, mainly based on PLM systems.
Therefore, we proposed a methodology for improving collaboration (internal and external) that consists in two processes: the first one consists on tracks generation and the second on collected tracks exploitation.

The evolving nature of the proposed approach is part of a process of quality and continuous improvement, enabling the optimization of business processes.
In addition, a reflexive posture may be conducted inducing an internal analysis of the proposed methodology in order to identify its limits.

At this stage, we identify a first limit regarding to the nature of collected information. An exchange and/or a dialog by email, this interaction is not observed on the server. Nevertheless this dialog may be an important element to explain the actor's activity (diversion processes for example) (Delattre M. & al 2007). Thus, a complementary observation source could be the use the navigator.
Moreover, the question of collected data relevance must be deepened on an experimental basis.

## LIST OF REFERENCES

*Journal articles:*

Delattre M. & Ocler R. 2007. Janus organisationnel : visages de l'homme dans l'organisation, *Revue économique et sociale*.

Marty J-C., Heraud J.M., Carron T., & France L. 2007. Matching the Performed Activity on an Educational Platform with a Recommended Pedagogical Scenario: a Multi Source Approach, *Journal of Interactive Learning Research*, vol 18, n° 2.

Mishra A. & Mishra D. 2006. Software quality assurance models in small and medium organisations: a comparison. *Int. J. of Information Technology and Management 2006* - Vol. 5, No.1 pp. 4-20.

*Book:*

Barrand J. 2006. Le manager agile, vers un nouveau management pour affronter la turbulence, Paris, Dunod.

Debaecker.D. 2004. PLM : La gestion collaborative du cycle de vie des produits, Lavoisier.

*Conference paper:*

Berthier D. 2006. Enrichissement de la modélisation des processus métiers par le paradigme des systèmes multi agents. *INT/GET*-2006.

Carron T., Marty J.C., Heraud J.M. & France L. 2006. Helping the teacher to re-organize tasks in a collaborative learning activity: an agent based approach, *ICALT*, Kerkrade, The Netherlands, 2006, p. 552-554.

Delattre M., 2001. L'activation du potentiel humain comme support fédérateur des projets de redressement d'une entreprise en difficulté : cas d'une entreprise de traitement de poissons, *Gestion 2000, n°5/2001* (septembre-octobre 2001), pp. 55 - 71.

El Kadiri S., Pernelle P., Delattre M. & Bouras A. 2008a. An approach to control collaborative processes in PLM systems, *Extended Product and Process Analysis aNd Design* – 20 & 21 March 2008, Bordeaux, France

El Kadiri S., Pernelle P., Delattre M., Bouras A. 2008b. Pilotage des processus collaboratifs dans les systèmes PLM. Quels indicateurs pour quelle évaluation des performances. *1$^{er}$ Congrès des innovations mécaniques CIM'08* – 28 & 29 avril 2008 Sousse, Tunisie

Gligor-Calin L. 2006. Aide à la compréhension du comportement de l'utilisateur par la transformation des traces collectées. *1ères Rencontres Jeunes Chercheurs en EIAH*, RJC-EIAH'2006, pages 88 à 95.

Green P. & Rosemann M. 2000. Integrated process modelling: an ontological evaluation, *Information Systems*, 25, 2, 73-87.

Heraud J-M., Marty J-C., France L. & Carron T. 2005. Une aide à l'interprétation de traces : application à l'amélioration de scénarios pédagogiques. *Environnements Informatiques pour l'Apprentissage Humain*, Montpellier 2005.

Ithnin N., Ibrahim O. & Wood-Harper T. 2006. Generic Collaboration Approach as a Strategy for the Implementation of an Electronic Government: A Case Study of Malaysia, *ICEG*-2006.

Morley C. 2004. Un cadre unificateur pour la représentation des processus, *Pre-ICIS*-2004.

Pol G., Merlo C., Legardeur J. & Jared G. 2007. Supporting collaboration in product design through PLM system customization. *International Conference on Product Lifecycle Management PLM'07*, PLM-SP3 – 2007, Proceedings pp. 21 - 30.